\documentclass[a4paper,fleqn]{cas-dc}

\usepackage[numbers,compress]{natbib}
\usepackage{footnote}
\usepackage{tablefootnote}
\usepackage{longtable}
\usepackage{threeparttablex}
\usepackage{caption}
\usepackage{siunitx}

\def\tsc#1{\csdef{#1}{\textsc{\lowercase{#1}}\xspace}}
\tsc{WGM}
\tsc{QE}
\tsc{EP}
\tsc{PMS}
\tsc{BEC}
\tsc{DE}

\begin{document}
\let\WriteBookmarks\relax
\def\floatpagepagefraction{1}
\def\textpagefraction{.001}

\shorttitle{Absolute intensity measurement of pulsed muon beams using in-beam activation}

\shortauthors{R. Mizuno et~al.}

\title[mode=title]{Absolute intensity measurement of pulsed muon beams using in-beam activation}  



%

\author[1,2,3]{R.~Mizuno}
\affiliation[1]{organization={CMMS, TRIUMF},
    city={Vancouver},
    postcode={BC V6T 2A3}, 
    country={Canada}}
\affiliation[2]{organization={Department of Physics, Graduate School of Science, The University of Tokyo},
    city={Bunkyo},
    postcode={Tokyo 113-0033}, 
    country={Japan}}
\cormark[1]
\fnmark[1]
\ead{rurie.mizuno@riken.jp}

\author[3]{M.~Niikura}
\author[3]{T.~Matsuzaki}
\affiliation[3]{organization={Nishina Center, RIKEN},
    city={Wako},
    postcode={Saitama 351-0198}, 
    country={Japan}}
    
\author[4]{A.~D.~Hillier}
\affiliation[4]{organization={ISIS Neutron and Muon Facility, STFC Rutherford Appleton Laboratory},
    city={Didcot},
    postcode={Oxfordshire OX11 0QX}, 
    country={United Kingdom}}

\author[5]{K.~Ishida}
\affiliation[5]{organization={High Energy Accelerator Research Organization (KEK)},
    city={Tsukuba},
    postcode={Ibaraki 305-0801}, 
    country={Japan}}

\author[6]{S.~Kawase}
\author[6]{T.~Kawata}
\author[6]{K.~Kitafuji}
\affiliation[6]{organization={Department of Advanced Energy Science and Engineering, Kyushu University},
    city={Kasuga},
    postcode={Fukuoka 816-8580}, 
    country={Japan}}

\author[7,5]{D.~Tomono}
\affiliation[7]{organization={Research Center for Nuclear Physics, The University of Osaka},
    city={Ibaraki},
    postcode={Osaka 567-0047}, 
    country={Japan}}



\begin{abstract}
The absolute number of negative muons contained in a beam is essential for many experiments at accelerator facilities, but determining this number in pulsed beams has been difficult, particularly at high intensities.
A method utilizing the yield of the $\beta$ delayed $\gamma$ rays from the residual nuclei following the muon nuclear capture reaction has recently been demonstrated for determining the muon number in the pulsed muon beam. 
In particular, the in-beam activation method employs isotopes with short lifetimes, enabling the beam intensity to be monitored during beam irradiation.
However, the applicability of this method was limited because the required production branching ratio (BR) data were available only for a limited number of isotopes, restricting its use to specific experimental conditions. 
To broaden the applicability of this method, we searched for new candidate isotopes that are suitable for beam intensity measurement using the in-beam activation method and assessed their feasibility to a wide range of experiments. 
Based on the strength of the production BR, the muon capture probability, the practical detection efficiency of the detector, and the rarity of the target material in the surrounding structures, the reactions $^\mathrm{nat}$Ag ($\mu^-, \nu_\mu x$) $^{107m}$Pd and $^\mathrm{nat}$Pd ($\mu^-, \nu_\mu x$) $^{107m, 105m}$Rh are identified as a useful references for determining the number of muons in pulsed beams using the in-beam activation method.


\end{abstract}



\maketitle

\section{Introduction} \label{sec:introduction}

The determination of beam intensity is a fundamental requirement for experiments at accelerator facilities, since the absolute beam intensity directly affects the normalization of cross-section measurements and the design of detector systems.
For pulsed muon beams, this determination is particularly challenging due to their characteristic time structure: a large number of muons arriving periodically within a short pulse width, typically on the order of tens of nanoseconds. 
The large number of muons concentrated within such a short pulse makes it difficult to convert the detector response of a beam monitor into an absolute number of muons per pulse.

A conventional approach to determine the pulsed muon beam intensity is direct counting using a scintillation counter.
Since the signal amplitude of the scintillator is proportional to the number of muons passing through it, the beam intensity can be estimated by measuring the single-muon response at low beam intensities and extrapolating to higher intensities.
This method provides a reasonable estimate when the number of muons per pulse is small, typically up to a few tens of muons per pulse~\cite{Niikura2024-ck, Sierawski2014-id}, but becomes unreliable in modern pulsed muon facilities such as J-PARC MUSE and RAL-ISIS, where thousands of muons arrive in a pulse.
Increasing the detector granularity can overcome this issue~\cite{Dal-Maso2023-om, Bonesini2019-bs}, but this requires a very high degree of segmentation and a large number of readout channels, and the calibration of each channel becomes increasingly demanding.
As an alternative to direct counting, the beam intensity has been estimated using secondary signals associated with muon decay and muonic X-ray emission.
The intensity of positive muons is typically inferred from the detection of decay positrons, while that of negative muons can be evaluated using muonic X-rays~\cite{Cook2017-cz, Zhang2026-qv}.
These methods are practical for monitoring beam intensity and relative variations.
However, their applicability to absolute intensity determination is limited by factors such as detection efficiency, energy thresholds, and pile-up effects at high intensities.
As a result, the absolute number of muons per pulse has not been determined with an accuracy better than approximately 10—20\%, especially for negative muons.

Recently, a method based on activation was proposed to determine the intensity of the negative muon beam~\cite{Mizuno2025-rt}.
Activation techniques enable such measurements by relating the number of radioactive nuclei produced in a target to the number of incident particles through known production probabilities.
In this approach, the production of radioactive nuclei following muon nuclear capture is utilized to determine the absolute beam intensity.
The in-beam activation method extends the conventional activation approach by measuring the decay radiation simultaneously with beam irradiation, enabling the measurement of short-lived radioactive nuclei with lifetimes down to a few tens of milliseconds~\cite{Niikura2024-ck}.
The application of this method to pulsed muon beams provides an ideal condition for decay measurements owing to the absence of beam-related background during the interpulse period, which is typically \SI{40}{ms} at J-PARC MUSE and \SI{20}{ms} at RAL-ISIS.
This allows the beam intensity to be determined promptly after irradiation, making online monitoring feasible.

The applicability of the in-beam activation method for absolute intensity measurement is currently limited by the availability of reliable absolute production branching ratio (BR) data.
Only a limited number of isotopes with well-established BRs exist, most of which have been determined using offline activation methods.
These measurements typically determine the BRs of long-lived isotopes and are therefore not well suited for in-beam activation.
In addition to the availability of BR data, the choice of target material is subject to several practical constraints.
The target must be sufficiently thick to fully stop the incoming muon beam, while background contributions from surrounding materials must be minimized.
These requirements favor materials that are readily available, mechanically stable, capable of being fabricated in large forms with sufficient thickness, and not commonly used as structural materials.
These considerations motivate the search for new candidate isotopes suitable for in-beam activation.
In this work, we measure the production BRs of several materials using the in-beam activation method and evaluate their suitability for absolute beam intensity determination.

The remainder of this paper is organized as follows: The method for determining the number of muons in pulsed beams using the activation method is described in Sec.~\ref{sec:method}. The experimental setup, analysis procedure, and results are presented in Secs.~\ref{sec:exp} and~\ref{sec:result}. 
The reliability of measured BRs are discussed in Sec.~\ref{sec:discussion-BR}, while their application to beam intensity measurements is discussed in Secs.~\ref{sec:discussion-Imuon} and \ref{sec:discussion-Imuon_practical}. Finally, Sec.~\ref{sec:conclusion} summarizes and concludes the paper.

\section{Method} \label{sec:method} 

In the activation-based method, the pulsed negative muon beam intensity is determined from the production rate of residual nuclei following muon nuclear capture. 
Negative muons stopped in a target form muonic atoms, and some of the muons are captured by protons in the nucleus via the elementary process $\mu^-+p\to n+\nu_\mu$. This muon nuclear capture reaction leaves the nucleus in a highly excited state, which subsequently deexcites by emitting $\gamma$ rays and several particles, resulting in radioactive residual nuclei with certain production probabilities.
The number of radioactive nuclei produced can be determined from $\beta$-delayed $\gamma$-ray measurements, particularly by detecting them during each interval between beam pulses in the in-beam activation method. 
Since the production yield of the residual nuclei is related to the number of incident negative muons, the total number of incident negative muons ($N_{\mu^-}$) can be derived from the $\beta$-delayed $\gamma$-ray yields ($N_\gamma$).

The relationship between $N_{\mu^-}$ and $N_\gamma$ is given by
\begin{equation}
    N_{\mu^-} = \frac{N_\gamma/I_\gamma \epsilon_\gamma}{b P_\mathrm{cap} r_\mathrm{decay}},
    \label{eq:Nmu}
\end{equation}
where $I_\gamma$ is the $\gamma$-ray emission probability per decay of the residual nucleus, $\epsilon_\gamma$ is the $\gamma$-ray detection efficiency, $b$ is the production BR of the residual nucleus, and $P_\mathrm{cap}$ is the muon capture probability. $r_\mathrm{decay}$ represents the ratio of the number of decays to the number of residual nuclei produced in the measurement time window, which is determined by the lifetime of the residual nucleus, the measurement time, and the beam intensity fluctuation. 
Using this equation, the absolute number of incident muons can be determined from the measured $\beta$-delayed $\gamma$-ray yield, because the other parameters in Eq.~(\ref{eq:Nmu}) can be determined independently, as described below.
This capability makes this in-beam activation method useful not only as a direct beam-intensity monitor but also as a calibration tool for other beam-monitor detectors. For example, when a plastic scintillator is operated simultaneously with the activation measurement, signals from the plastic scintillator can be calibrated to the absolute number of incident muons determined from Eq.~(\ref{eq:Nmu}). 
Such calibration is particularly valuable when pulse-by-pulse beam monitoring is required.
Direct beam-intensity monitoring using in-beam activation requires the beam to be fully stopped in the reference sample, whereas the use of other beam-monitor detectors, which are calibrated using the in-beam activation, enables simultaneous beam monitoring without consuming the beam.

The parameters in Eq.~(\ref{eq:Nmu}) are obtained from evaluated nuclear data or determined independently for each experiment.
The $\gamma$-ray emission probability $I_\gamma$ can be obtained from nuclear data libraries such as Evaluated Nuclear Structure Data File (ENSDF)~\cite{ENSDF} and Decay Data Evaluation Project (DDEP)~\cite{DDEP, Kellett2017}.
The detection efficiency $\epsilon_\gamma$ depends on each experimental setup.
The muon capture probability $P_{\mathrm{cap}}$ can be derived from the lifetime of the muonic atom ($\tau$) as
\begin{equation}
    P_\mathrm{cap} = 1 - \frac{\tau}{\tau_{\mu^+}}Q,
\end{equation}
where $\tau_{\mu^+}$ is the lifetime of the positive muon, $\tau_{\mu^+} = \SI{2.196811(22)}{\micro \second}$, and $Q$ is the Huff factor~\cite{Uesaka2026-kd}, which accounts for the increase in the partial lifetime of the decay-in-orbit of the muonic atom ($\tau_{\mu^+}/Q$ where $Q<1$) due to the muon’s binding to an atomic orbital.

$r_\mathrm{decay}$ is calculated depending on the measuring isotopes.
When assuming a constant beam intensity ($I_\mathrm{beam}$), $r_\mathrm{decay}$ can be expressed as
\begin{equation}
    r_\mathrm{decay} = \frac{
        \int^{t_\mathrm{stop}}_{t_\mathrm{start}}
        I_\mathrm{beam}
        \left[ 1-\exp(-\lambda (t-t_\mathrm{start})) \right] dt
    }
    {I_\mathrm{beam}(t_\mathrm{stop} - t_\mathrm{start})},
\label{eq:P_decay}
\end{equation}
where $\lambda$ is the decay constant of the residual nucleus ($\lambda = \ln(2)/T_{1/2}$), $t_\mathrm{start}$ is the start time of the irradiation, and $t_\mathrm{stop}$ denotes the stop time of the measurement. It is assumed that no residual nuclei are present at the beginning of the irradiation, and the measurement stops during the irradiation.
The value of $r_\mathrm{decay}$ approaches one when the irradiation time is long enough compared with the lifetime of a given residual nucleus, whereas it becomes a small value when the irradiation time is much shorter than the lifetime.
Among the parameters in Eq.~(\ref{eq:Nmu}), the production BR ($b$) is currently the least well known.
Although production BRs have been measured for some isotopes using continuous beams or low-intensity pulsed beams under countable conditions, the available data remain sparse. Consequently, the lack of reliable BR data currently limits the applicability of the in-beam activation method for absolute beam-intensity determination.

To increase the number of available $b$ and search for suitable target materials for this method, the production BRs of several candidate targets were measured using the pulsed muon beam under the countable condition in this study.
Since the activation measurement relies on $\beta$-delayed $\gamma$ rays emitted during irradiation, the main residual nuclei should have relatively short lifetimes so that a sufficiently large fraction of decays occurs during the measurement. 
The optimal lifetime depends on the intended application. Short-lived nuclei (typically with half-lives shorter than about a few minutes) are preferred for online beam monitoring because they respond quickly to beam fluctuations, whereas nuclei with lifetimes of several tens of minutes can still be suitable when the objective is to determine the total number of incident muons.

Based on these requirements, together with the practical constraints described in Sec.~\ref{sec:introduction}, natural abundance samples of Cu, Zn, and Ag were selected as candidate targets. 
These elements were chosen because intense $\beta$-delayed $\gamma$ rays are expected from residual isotopes with lifetimes of shorter than a few minutes. 
All of these samples are easy to fabricate into the target geometry and are not commonly used in structural materials. In addition, Cu and Ag have only two isotopes in nature, making it relatively easy to discuss the reliability of the results. 

\section{Experiment and Analysis} \label{sec:exp}

Measurement of the production BRs following muon nuclear capture was conducted at Rutherford Appleton Laboratory (RAL).
The experimental setup, analysis procedure, and notation are the same as those used in our previous in-beam activation measurement for Al and Si isotopes~\cite{Mizuno2025-rt}.

The experiment was performed at Port 4 of the ISIS Neutron and Muon Source (RIKEN-RAL beamline)~\cite{Matsuzaki2001-pg, Hillier2018-vd}.
A proton beam was accelerated to \SI{800}{MeV} by the ISIS rapid cycling synchrotron, and four out of every five pulses were directed to Target Station 1 (TS1).
Pions were produced by irradiating the proton beam on a graphite target at TS1, and negative muons produced from pion decay were then transported to Port 4.

\begin{figure}
    \centering
    \includegraphics[width=0.99\linewidth]{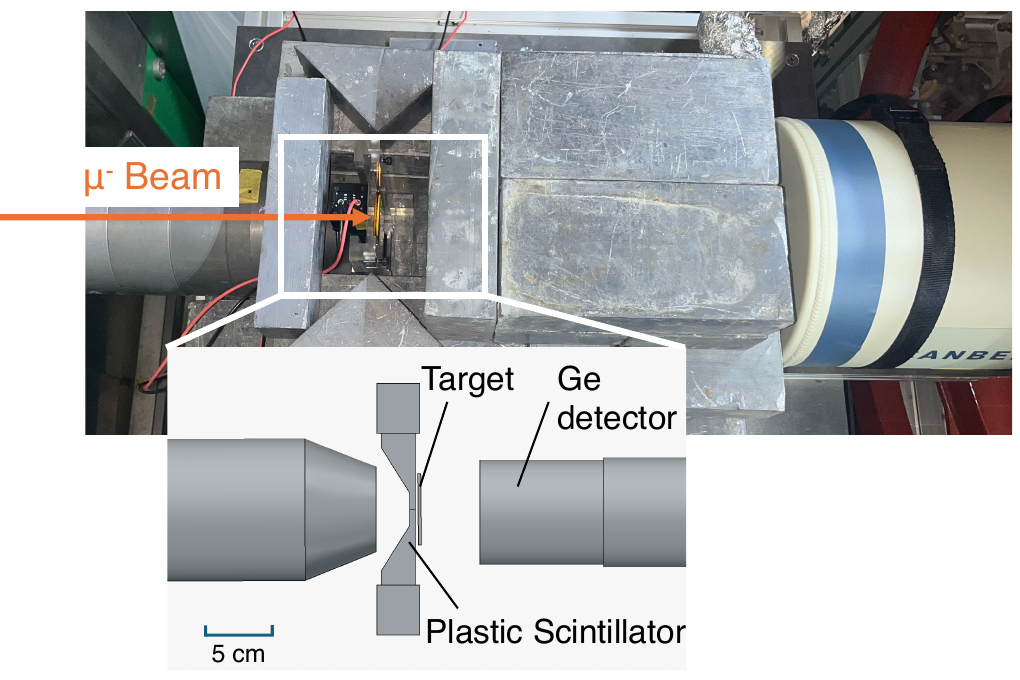}
    \caption{Photograph of the detector setup at Port 4 of the RIKEN-RAL muon facility~\cite{Mizuno2026-yp}. The plastic scintillator is placed upstream of the target, and the germanium detector is positioned at \SI{0}{\degree} with respect to the beam direction. Enlarged view is not to scale.}
    \label{fig:setup}
\end{figure}

\begin{table*}[width=1.5\linewidth]
\begin{threeparttable}
    \centering \small
    \caption{Summary of the materials and geometries of the targets. The muon capture probability ($P_\mathrm{cap}$) is listed in the fifth column~\cite{Suzuki1987-nk, Mizuno2025-wx}. The beam irradiation and decay measurement times are given in the last two columns.}
    \label{tab:targets}
    \begin{tabular}{ccccccc}\toprule
    Target    & Form  & Size(mm$^3$) & Purity & $P_\mathrm{cap}$(\%) & Irradiation & Decay \\\midrule
    $^\mathrm{nat}$Cu\footnotemark[1]  & Metal plate   & 50$\times$50$\times$0.6  &  99.9$+$\% & 92.8(6)\cite{Suzuki1987-nk} & 10 h & 2.0 h  \\ 
    $^\mathrm{nat}$Zn\footnotemark[2]      & Metal plate   & 50$\times$55$\times$1.0   & 99.2\% & 93.0(6)\cite{Suzuki1987-nk} & 5.0 h  & 1.0 h   \\
    $^\mathrm{nat}$Ag\footnotemark[3]  & Metal plate  & 50$\times$50$\times$0.75   & 99.99+\% & 96.3(6)\cite{Mizuno2025-wx} & 0.83 h & 0.08 h  \\
    Empty      & Acrylic case & - & - & - & 4.0h  & 0.0h \\\bottomrule
    \end{tabular}
    \begin{tablenotes}\footnotesize
    \item[1]{The natural isotopic abundance of $^\mathrm{nat}$Cu is $^{63}$Cu: 69.15(15)\% and $^{65}$Cu: 30.85(15)\%~\cite{Meija2016-vz}.}
    \item[2]{The natural isotopic abundance of $^\mathrm{nat}$Zn is $^{64}$Zn: 49.17(75)\%, $^{66}$Zn: 27.73(98)\%, $^{67}$Zn: 4.04(16)\%, $^{68}$Zn: 18.45(63)\%, and $^{70}$Zn: 0.61(10)\%~\cite{Meija2016-vz}.}
    \item[3]{The natural isotopic abundance of $^\mathrm{nat}$Ag is $^{107}$Ag: 51.839(8)\% and $^{109}$Ag: 48.161(8)\%~\cite{Meija2016-vz}.}
    \end{tablenotes}
\end{threeparttable}
\end{table*}

The experimental setup is shown in Fig.~\ref{fig:setup}.
The momentum of the muon beam ($p$) was \SI{36}{MeV/c} with the rms spread ($\Delta p/p$) of 4\%.
The number of muons irradiating the targets was measured using a plastic scintillator (EJ-212, Eljen) with a thickness of \SI{0.5}{mm} placed \SI{35}{mm} downstream of the end of the beam collimator. The scintillation light was read out by two photomultipliers (H11934-100-10, Hamamatsu).
The absolute number of muons per pulse was deduced from the charge integral of the plastic scintillator signal, which was calibrated using the known production BR of the $^{27}$Al($\mu^-$, $\nu_\mu$)$^{27}$Mg reaction, 9.90(33)\%~\cite{Mizuno2025-rt}.
The beam intensity was approximately 40 particles per pulse.
The targets were installed \SI{1}{mm} downstream of the plastic scintillator.
Table~\ref{tab:targets} summarizes the materials and geometries of the targets.
Natural-abundance Cu, Zn, and Ag plates, denoted as $^\mathrm{nat}$Cu, $^\mathrm{nat}$Zn, $^\mathrm{nat}$Ag, respectively, were irradiated with muons.
The target thicknesses were chosen to ensure that the muons were fully stopped in the target.
A blank target consisting of an empty acrylic case was also measured to evaluate the background.

The $\beta$-delayed $\gamma$ rays were measured using a germanium detector (Canberra GX5019)~\cite{Mizuno2024-br} placed \SI{58}{mm} downstream of the target.
To reduce environmental background, the plastic scintillator, target, and detector were enclosed in lead shielding.
The energy and efficiency of the germanium detector were calibrated using standard $\gamma$-ray sources, $^{60}$Co, $^{133}$Ba, and $^{137}$Cs.
The efficiency was corrected using Geant4 simulations~\cite{Agostinelli2003-eh,Allison2006-ho,Allison2016-zr} to account for self-absorption effects in each target. 
The shift of the energy peaks during the measurement was evaluated using the reference energy peaks, assuming a linear gain drift.
The energy peaks at 139.68 ($^{75m}$Ge), 198.39 ($^{71m}$Ge), 511, 1063.66 ($^{207m}$Ge) keV were used as the reference energies.

Both the plastic scintillator and germanium detector signals were acquired using two waveform digitizer boards, CAEN V1730B and V1730SD~\cite{CAENV1730}. 
For the plastic scintillator, all waveforms were recorded triggered by a 50~Hz synchrotron signal. The energy and timestamp data of the germanium detector were analyzed and recorded under self-trigger conditions using the digital pulse-processing firmware for pulse height analysis (DPP-PHA, CAEN).


\begin{figure}
    \centering
    \includegraphics[width=0.9\linewidth]{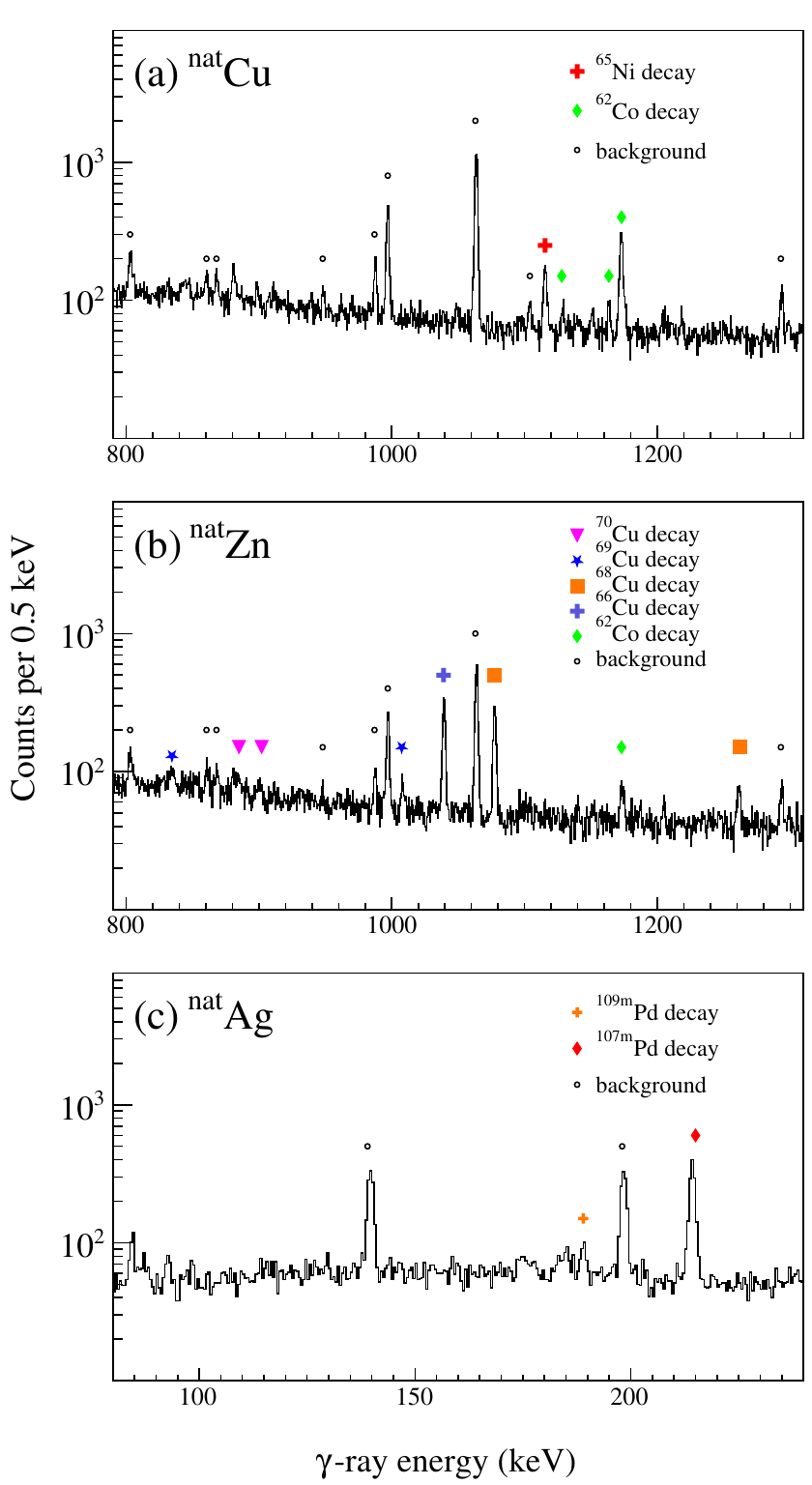}
    \caption{Gamma-ray energy spectra obtained with (a) $^\mathrm{nat}$Cu, (b) $^\mathrm{nat}$Zn, and (c) $^\mathrm{nat}$Ag targets.}
    \label{fig:energy}
\end{figure}

The production BR was deduced according to the procedure described in previous studies~\cite{Niikura2024-ck, Mizuno2025-rt}.
Figure~\ref{fig:energy} shows the energy spectra obtained with the germanium detector.
Since multiple muons simultaneously form muonic atoms and subsequently emit several muonic X-rays during approximately a 150-ns beam pulse, the germanium detector signals are saturated for an order of milliseconds after beam irradiation. The saturation period was evaluated to be 0.17 ms and treated as the analysis dead time.
$\beta$-delayed $\gamma$ rays were identified by their peak energy in the spectrum after excluding data during the above dead time.
To account for any overlaps with the background components, spectra obtained in the empty target measurement and background measurement were subtracted for each energy peak, and the existence of single- and double-escapes from higher energy peaks was examined. 
The production BRs were determined from the measured muon numbers and the $\gamma$-ray yield using Eq.~(\ref{eq:Nmu}) because the number of muons was measured independently using the plastic scintillator in this study. 
The production BR is defined as the independent production yield of a ground or isomeric state normalized by the number of muon nuclear captures. 
The independent yield accounts only for direct production, excluding contributions from secondary processes such as $\beta$-decay or isomeric decay of other nuclei that subsequently populate the same state.
For the nuclei that are potentially produced by decay from isomeric states, $^{62}$Co and $^{70}$Cu in this study, the production number from the decay was evaluated and subtracted. 
Some of the observed nuclei could also potentially be populated by decay from other nuclei.
However, no such feeding nuclei were observed in the present study, and their contributions were therefore not included.

\section{Results} \label{sec:result}

\begin{table*}[htbp]
\centering
\begin{threeparttable}
\caption{
Absolute production BRs of each nucleus produced following muon nuclear capture of $^\mathrm{nat}$Cu, $^\mathrm{nat}$Zn, and $^\mathrm{nat}$Ag.
The muon capture probabilities, $P_\mathrm{cap}$, listed in Table~\ref{tab:targets} are used.
The target nuclei (Target), Residual nuclei (Residual), spin-parity of the residual states (State), decay modes (Decay), half-lives ($T_{1/2}$), $\gamma$-ray energies ($E_\gamma$) and intensities ($I_\gamma$) are listed. These decay properties are obtained from ENSDF~\cite{Singh2025-pz, Chen2025-xt, Browne2010-fs, Junde2005-zj, McCutchan2012-oe,Nesaraja2014-qu, Gurdal2016-ot, Blachot2008-fs, Kumar2016-lq}.
BRs determined using each $\gamma$-ray intensity ($b_\gamma$) and the weighted average values for each nucleus ($b$) are presented in the last two columns.
}
\label{tab:Result}

\begin{tabular}{cccccccll}\toprule
Target & Residual   & State & Decay & $T_{1/2}$ & $E_\gamma$ (keV) & $I_\gamma$ (\%) & \multicolumn{1}{c}{$b_\gamma$ (\%)\footnotemark[1]} & \multicolumn{1}{c}{$b$ (\%)\footnotemark[1]}\\ \midrule
    
    $^\mathrm{nat}$Cu 
    & $^{65}$Ni & 5/2$^-$ & $\beta^-$ & 2.5175(5) h & 366.27  & 4.753(35)  & 1.60(31) \\
    &         &         &           &               & 1115.53 & 15.522(47) & 1.35(10) \\
    &         &         &           &               & 1481.84	& 23.59(24)\footnotemark[2] & 1.41(8) \\
    &         &     & & & & $\Delta I_\gamma^{\mathrm{abs}}/I_\gamma^{\mathrm{abs}}$=0.59\% &  & 1.40(6)(5)\footnotemark[10]\\
   &$^{62}$Co & 2$^+$   & $\beta^-$ & 1.54(10) min & 1128.9 & 10.3(5) & 0.46(12)\footnotemark[3]\\
    &         &         &           &              & 1172.9 & 83.2(40)\footnotemark[2] & 0.520(38)\footnotemark[3] \\
    &         &     & & & & $\Delta I_\gamma^{\mathrm{abs}}/I_\gamma^{\mathrm{abs}}$=0.48\%  & & 0.515(49)(19)\footnotemark[4]\\
   &$^{62m}$Co& 5$^+$   & $\beta^-$ & 13.86(9) min & 1163.5 & 68.0(8) & 0.103(14)\\
    &         &     & & & & $\Delta I_\gamma^{\mathrm{abs}}/I_\gamma^{\mathrm{abs}}$=0.51\%  & & 0.103(14)(4)\\
    \hline
    \addlinespace
    $^\mathrm{nat}$Zn 
    &$^{70}$Cu & 6$^-$ & $\beta^-$ & 44.5(2) s & 884.88  & 100.00(40)  & 0.041(13)\footnotemark[7] \\
    &          &       &           &           & 901.7   & 99.70(40)   & 0.040(17)\footnotemark[7] \\
    &          &     & & & & $\Delta I_\gamma^{\mathrm{abs}}/I_\gamma^{\mathrm{abs}}$=4.0\% &  & $<$0.041(10)(1)\footnotemark[7]\footnotemark[10]\\
  &$^{70m1}$Cu & 3$^-$ & $\beta^-$ & 33(2) s & 884.88  & 49.0(25)  & 0.042(14)\footnotemark[7] \\
    &          &       &           &         & 901.7   & 25.0(25)  & 0.055(24)\footnotemark[7] \\
    &          &     & & & & $\Delta I_\gamma^{\mathrm{abs}}/I_\gamma^{\mathrm{abs}}$=18\% &  & $<$0.045(15)(2)\footnotemark[7]\\
  &$^{70m2}$Cu & 1$^+$ & $\beta^-$ & 6.6(2) s & 884.88  & 88.4(18)\footnotemark[6]  & 0.044(14)\footnotemark[7] \\
    &          &     & & & & $\Delta I_\gamma^{\mathrm{abs}}/I_\gamma^{\mathrm{abs}}$=incl.\footnotemark[5] & & $<$0.034\footnotemark[7] \\
    &$^{69}$Cu & 3/2$^-$ & $\beta^-$ & 2.85(15) min & 834.4 & 13.10(47)\footnotemark[8] & 0.38(11) & \\
    &          &         &           &              & 1007.5 & 23.4(7)\footnotemark[8]  & 0.39(9) & \\
    &          &     & & & & $\Delta I_\gamma^{\mathrm{abs}}/I_\gamma^{\mathrm{abs}}$=16\%\footnotemark[8] &  & 0.39(9)(1)\footnotemark[10]\\
    &$^{68}$Cu & 1$^+$   & $\beta^-$ & 30.9(6) s  & 1077.7 & 61.1(35) & 1.39(9) & \\
    &          &         &           &            & 1261.8 & 14.5(10) & 1.25(18) \\
    &          &     & & & & $\Delta I_\gamma^{\mathrm{abs}}/I_\gamma^{\mathrm{abs}}$=1.7\%  & & 1.36(9)(5) \\
    &$^{67}$Cu & 3/2$^-$ & $\beta^-$ & 61.83(12) h & 93.31  & 16.10(20) & 13.2(27) & \\
    &          &         &           &             & 184.58 & 48.70(30) & 11.8(10) \\
    &          &     & & & & $\Delta I_\gamma^{\mathrm{abs}}/I_\gamma^{\mathrm{abs}}$=incl.\footnotemark[5] &  & 11.9(10)(4)\\
    &$^{66}$Cu & 1$^+$   & $\beta^-$ & 5.120(14) min  & 1039.2 & 9.23(9) & 9.30(32) & \\
    &          &     & & & & $\Delta I_\gamma^{\mathrm{abs}}/I_\gamma^{\mathrm{abs}}$=incl.\footnotemark[5]  & & 9.30(32)(34)\\
    &$^{62}$Cu & 1$^+$   & $\epsilon$+$\beta^+$ & 9.670(8) min  & 875.66 & 0.150(7) & - & $<$13.5 \\
    &          &     & & & & $\Delta I_\gamma^{\mathrm{abs}}/I_\gamma^{\mathrm{abs}}$=5\%  & & \\
    &$^{62}$Co & 2$^+$   & $\beta^-$ & 1.54(10) min  & 1172.9 & 83.2(40)\footnotemark[2] & $<$0.078(24) & \\
    &          &     & & & & $\Delta I_\gamma^{\mathrm{abs}}/I_\gamma^{\mathrm{abs}}$=incl.\footnotemark[5]  & & $<$0.078(25)(3)\footnotemark[9]\\
    &$^{62m}$Co& 5$^+$   & $\beta^-$ & 13.86(9) min  & 1163.5 & 68.0(8) & - & $<$ 0.051\\
    &          &     & & & & $\Delta I_\gamma^{\mathrm{abs}}/I_\gamma^{\mathrm{abs}}$=0.51\%  & & \\
    \bottomrule
\end{tabular}
\end{threeparttable}
\end{table*}
\begin{table*}[htbp]\ContinuedFloat
\centering
\caption*{(continued.)}
\begin{threeparttable}
\begin{tabular}{cccccccll}\toprule
Target & Residual   & State & Decay & $T_{1/2}$ & $E_\gamma$ (keV) & $I_\gamma$ (\%) & \multicolumn{1}{c}{$b_\gamma$ (\%)\footnotemark[1]} & \multicolumn{1}{c}{$b$ (\%)\footnotemark[1]}\\ \midrule
    $^\mathrm{nat}$Ag
    &$^{109m}$Pd & 11/2$^-$ & IT & 4.703(9) min & 188.9 & 56.00(30)  & 0.225(48) \\
    &          &     & & & & $\Delta I_\gamma^{\mathrm{abs}}/I_\gamma^{\mathrm{abs}}$=Incl.\footnotemark[5] &  & 0.225(48)(8)\footnotemark[10]\\
    &$^{107m}$Pd & 11/2$^-$ & IT & 21.3(5) s & 214.9 & 68.7(7)  & 1.79(7) \\
    &          &     & & & & $\Delta I_\gamma^{\mathrm{abs}}/I_\gamma^{\mathrm{abs}}$=Incl.\footnotemark[5] &  & 1.79(7)(7)\\
    \bottomrule    
\end{tabular}
\begin{tablenotes}\scriptsize
\item[1]{The uncertainties in the $b_\gamma$ column include the statistical uncertainty from the $\gamma$-ray peak count and relative uncertainty from $I_\gamma$ ($I_\gamma^{rel}$). The first set of parentheses in the weighted average row ($b$) indicates relative uncertainty, encompassing the statistical uncertainty, uncertainties from $I_\gamma$ (relative and absolute), and the lifetime. The second set of parentheses includes uncertainties in the efficiency of the germanium detector, muon capture rate, and number of muons counted with the plastic scintillator, excluding the relative uncertainty. The uncertainties from the lifetime and muon capture rate were negligible compared with others.}
    \item[2]{$\Delta I_\gamma^{rel}$ of these peaks are not provided in the ENSDF database and were estimated from other $\Delta I_\gamma^{rel}$.}
    \item[3]{The peaks at 1128.9 and 1172.9 keV originated both from $^{62m}$Co and $^{62}$Co. The amount originating from $^{62m}$Co was calculated from the BR of $^{62m}$Co and subtracted.}
    \item[4]{The lower limit of $^{62}$Co is 0.46(4)\% calculated from the detection limit of $^{62}$Fe (0.058\%).} 
    \item[5]{$I_\gamma$ uncertainty includes both $\Delta I_\gamma^{rel}$ and $\Delta I_\gamma^{abs}$ for the nuclei denoted as $\Delta I_\gamma^{abs}/I_\gamma^{abs}$=incl.}
    \item[6]{The $\gamma$ ray intensity was provided as the upper limit.}
    \item[7]{The energy peaks at 884.88 and 901.7 keV originate from $^{70}$Cu and its isomers. The values in the column of $b_\gamma$ are derived by neglecting the direct feed to the other isomer or ground state. 
    The upper limits of $^{70, 70m1}$Cu shown in the column of $b$ were determined with the same assumption. 
    The upper limit of $^{70m2}$Cu (1$^+$) was calculated by the distribution of the yields of energy peaks. See details in the text.}
    \item[8]{The latest evaluation for $^{69}$Cu decay~\cite{Nesaraja2026-qz} reports that the $I_\gamma$ of 834.4 keV is 5.6(2)\%, that of 1007.5 keV is 10.0(3)\%, and $\Delta I_\gamma^{abs}/I_\gamma^{abs}$=20\%. Considering the large deviation from the previously evaluated value, we adopted the value in Ref.~\cite{Nesaraja2014-qu}.}
    \item[9]{The lower limit of $^{62}$Co is 0.023\%, assuming the presence of $^{62}$Cu at its upper limit and neglecting the direct transition to $^{62m}$Co.}
    \item[10]{For these nuclei, the production channel can be uniquely identified owing to their neutron excess because muon nuclear capture makes the nucleus more neutron-rich. 
    The BR of the $^{65}\mathrm{Cu} (\mu^-, \nu_\mu) ^{65}\mathrm{Ni}$ reaction is 4.53(26)\%, BRs of $^{70}\mathrm{Zn} (\mu^-, \nu_\mu n) ^{69}\mathrm{Cu}$ and $^{70}\mathrm{Zn} (\mu^-, \nu_\mu) ^{70}\mathrm{Cu}$ are 63(15)\% and 6.7(17)\%, respectively, where the latter assumes no population to the isomeric state. The BR of $^{109}\mathrm{Ag} (\mu^-, \nu_\mu) ^{109m}\mathrm{Pd}$ is 0.47(10)\%.}

\end{tablenotes}
\end{threeparttable}
\end{table*}

Table~\ref{tab:Result} lists the measured absolute BRs of nuclei produced following the muon nuclear capture reaction.
For each transition, the target nucleus, residual nucleus, its spin-parity, decay mode, half-life ($T_{1/2}$), $\gamma$-ray energy ($E_\gamma$), and intensity ($I_\gamma$) are adopted from ENSDF~\cite{ENSDF} and tabulated in the table.
BRs derived from each measured $\gamma$-ray ($b_\gamma$) are given for each $\gamma$-ray energy, and the weighted average BR for each residual nucleus ($b$) is also listed.
For some isotopes, the upper limit of population from the decay of non-detected nuclei was evaluated, and the corresponding lower limits are given in the footnote of the table.

The uncertainties of $b_\gamma$ include statistical uncertainty of the photo-peak counts and relative uncertainty in $I_\gamma$.
The uncertainties in the columns of weighted average values ($b$) are separated into two components, one independent for each isotope and the other common to all detected isotopes in each target.
The first parentheses represent the relative uncertainty, including statistical uncertainty, uncertainties from $I_\gamma$ (both relative and absolute), and the half-life of the nuclei.
The second parentheses include the uncertainties in the photo-peak detection efficiency of the germanium detector, muon capture probability, and the number of muons irradiating the target while excluding the relative uncertainty.

The energies of $\beta$-delayed $\gamma$ rays from the ground state of $^{62}$Co at 1129 and 1173 keV overlap with those of $^{62m}$Co, and the statistics originating from the ground state decay were evaluated by subtracting the contribution from the decay of the isomeric state (table footnote~3).
In the analysis of the Cu target, the production BR of the isomeric state was evaluated using the energy peak at 1163.5 keV, which is unique to the isomeric state.

\begin{figure}
    \centering
    \includegraphics[width=0.9\linewidth]{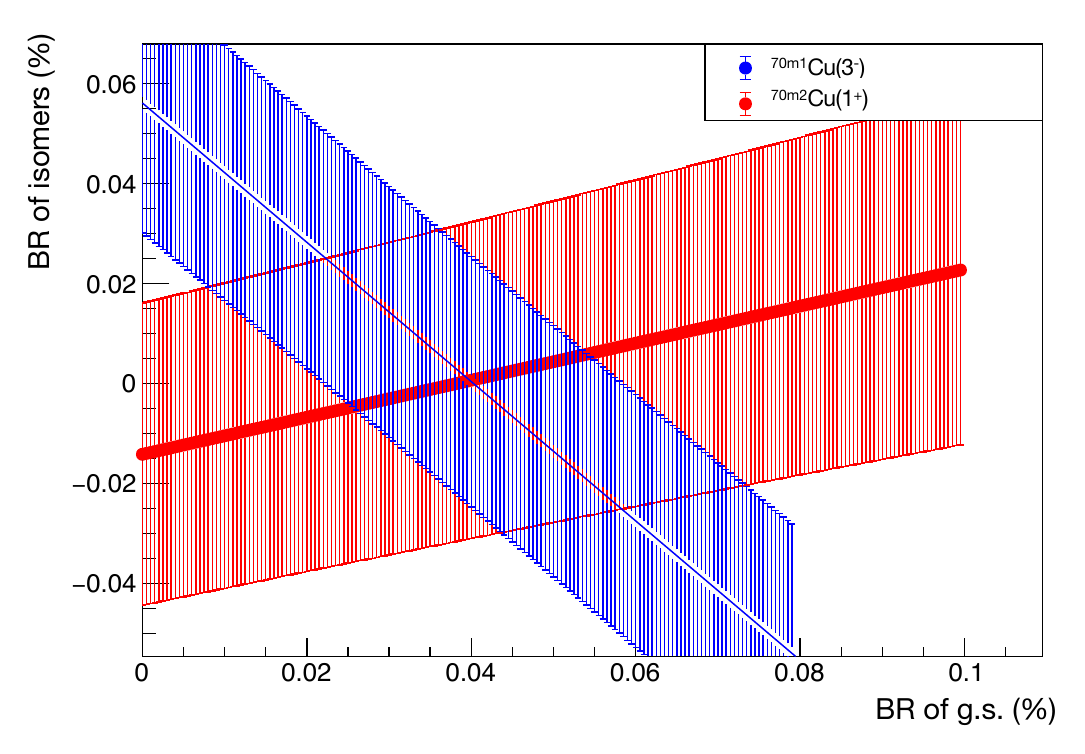}
    \caption{BRs of the isomeric and ground state of $^{70}$Cu observed using $^\mathrm{nat}$Zn target. The blue and red lines correspond to the BRs of isomeric states (blue: $^{70m1}$Cu($3^-$), red: $^{70m2}$Cu($1^+$)) when the BR of the ground state is fixed on the value shown on the horizontal axis.
    }
    \label{fig:70Cu_BR}
\end{figure}

On the other hand, $^{62, 62m}$Co and $^{62}$Cu are candidates for the origin of the energy peak at 1173 keV observed with $^\mathrm{nat}$Zn target. The upper limits of the production BR of $^{62m}$Co and $^{62}$Cu were calculated based on the non-observation of the corresponding energy peak as listed in Table~\ref{tab:Result}. 
The upper limit of the production BR of $^{62}$Co was calculated assuming all the statistics of the 1173 keV originated from $^{62}$Co.
When evaluating the statistics of the energy peak at 834.4 keV from $^{69}$Cu, the energy peak at 833 keV from $^{66}$Cu was included in the fitting process. The amount of the 833 keV component was calculated from the production BR of $^{66}$Cu.
The production BRs of $^{70, 70m1, 70m2}$Cu from the $^\mathrm{nat}$Zn target (labeled with table footnote 7) were evaluated with the following procedure.
$^{70}$Cu has two isomeric states, and both isomeric states decay via internal transition and $\beta$-decay.
Two energy peaks at 884.9 and 901.7 keV were observed that correspond to the $\beta$-decay from $^{70, 70m1, 70m2}$Cu, and no energy peak unique to a specific state was observed.
The direct production BRs for each state were calculated so as to reproduce the statistics of these energy peaks, while accounting for population from isomeric states through internal transitions. 
Once the BR of the ground state is fixed, the production BRs of isomeric states are uniquely determined from the observed $\gamma$-ray yields.
Figure~\ref{fig:70Cu_BR} shows the possible region of production BR. 
The blue and red lines correspond to the BR of isomeric states (blue: 3$^-$, red: 1$^+$) as a function of the ground state production BR (horizontal axis).
The allowed region of the production BR was evaluated so that all states have positive BR values within the uncertainty region.
The upper limits of the production BRs of the ground state and the first isomeric state ($^{70m1}$Cu) were determined under the assumption that all detected yields originate from each state, respectively, since this condition satisfies the above requirement for all BR values within the uncertainties.
The upper limit of the second isomeric state ($^{70m2}$Cu) was calculated using the upper limit of the BR of the ground state to satisfy the same requirement. 

For $^{109m, 107m}$Pd observed using $^\mathrm{nat}$Ag target, there is no candidate for the contamination of the decay component or possible background.

\section{Discussion}

The absolute production BRs for $^\mathrm{nat}$Cu, $^\mathrm{nat}$Zn, and $^\mathrm{nat}$Ag were obtained.
This section first discusses the general trends in the measured BRs by comparisons with previous results and model calculations.
Then, their applicability to beam intensity measurements is examined, with particular emphasis on identifying suitable target materials for this purpose in Sec.~\ref{sec:discussion-Imuon}.
Finally, a practical use of this in-beam activation method for beam intensity measurement will be presented in Sec.~\ref{sec:discussion-Imuon_practical}.
Hereafter, the reaction channels following muon capture are denoted by the number of emitted protons and neutrons; for example, $0p0n$ denotes the $(\mu^-,\nu_\mu)$ channel and $1p2n$ denotes the $(\mu^-, \nu_\mu\,1p2n)$ channel.

\subsection{Production Branching Ratio} \label{sec:discussion-BR}
All observed BRs are roughly consistent with the general trend reported previously~\cite{Niikura2024-ck, Mizuno2025-rt, Measday2001-mw, Yamaguchi2025-jn}: neutron emission ($0pxn$ channels) is dominant and charged particle emission is less probable.
The results of $^\mathrm{nat}$Zn have implications for the dominance of neutron emission channels.
The sum of the production BRs of $^{67,68,69}$Cu is 13.7(11)\%, which is attributed to neutron emissions from muonic $^{67,68,70}$Zn. 
Since these isotopes account for only 23.1\% of the natural abundance in total, the production of $^{67,68,69}$Cu is dominant among them.
In particular, the BR of $^{69}$Cu, which originates only from $^{70}$Zn, is 0.39(9)\% and results in a large BR of $0p1n$ channel from $^{70}$Zn by considering the small fraction of $^{70}$Zn among natural abundance (0.61\%). 
For charged particle emission channels, only $1pxn$ channels of $^\mathrm{nat}$Cu were observed with small production BR: 0.515(53)\% and 0.103(15)\% for $^{62}$Co and $^{62m}$Co, respectively. 
This is consistent with the trend reported by Wyttenbach et al.~\cite{Wyttenbach1978-ks} that charged particle emission is suppressed by the Coulomb barrier, which becomes more effective with increasing atomic number.

\begin{table}
    \begin{threeparttable}
    \centering\small
    \caption{
        Comparison between the present and previous values.
        All previous values were measured using the activation method.
    }
    \label{tab:previous}
    \begin{tabular*}{\tblwidth}{@{} CLLL @{}} 
    \toprule
    Reaction & Present (\%) & Previous (\%)\\ \midrule
    $^\mathrm{nat}\mathrm{Cu}(\mu^-, \nu_\mu){}^{65}\mathrm{Ni}$ & 1.40(8) & 1.79(9)\footnotemark[1]~\cite{Wyttenbach-in-Mukhopadhyay}\\
    && 1.91\footnotemark[2]~\cite{Heisinger2002-nm}\\ 
    $^\mathrm{nat}$Cu ($\mu^-, \nu_\mu x$) $^{62}$Co & 0.618(56) & 0.63(10)~\cite{Wyttenbach1978-ks}\\ 
    $^\mathrm{nat}$Cu ($\mu^-, \nu_\mu x$) $^{62m}$Co & 0.103(15) & 0.16(1)~\cite{Heisinger2002-nm}\\ 
    \bottomrule
    \end{tabular*}
    \begin{tablenotes}\footnotesize
    \item[]{Uncertainties include statistical and systematic uncertainties.}
    \item[1]{Derived from the BR of 5.8(3)\% for $^{65}\mathrm{Cu}(\mu^-, \nu_\mu){}^{65}\mathrm{Ni}$ in Ref.~\cite{Wyttenbach-in-Mukhopadhyay}.}
    \item[2]{Uncertainty was reported as less than 0.1\% in the original paper.}
    \end{tablenotes}
    \end{threeparttable}
\end{table}

Several previous studies have measured the production BRs following muon nuclear capture on $^\mathrm{nat}$Cu using the activation method~\cite{Wyttenbach-in-Mukhopadhyay,Heisinger2002-nm,Wyttenbach1978-ks}.
Table~\ref{tab:previous} compares the present BRs with previous results.
The BR of $^\mathrm{nat}$Cu$(\mu^-, \nu_\mu x)^{62}$Co, which is the sum of the ground state and isomeric state BRs, is consistent with the previous value~\cite{Wyttenbach1978-ks}, whereas the BRs of $^{62m}$Co and $^{65}$Ni show systematically larger values in the previous measurements~\cite{Heisinger2002-nm, Wyttenbach-in-Mukhopadhyay}.
The overestimation of the production BR in Ref.~\cite{Heisinger2002-nm} was also reported in our previous study for Al and Si isotopes~\cite{Mizuno2025-rt}.
The cause of the discrepancy in the $^{65}$Cu($\mu^-, \nu_\mu$)$^{65}$Ni reaction is uncertain due to insufficient experimental details in the original publication.
These discrepancies with previous measurements are not included as additional uncertainties in the subsequent discussion of the muon beam intensity monitoring method because the limited information on the absolute normalization of the previously reported BRs makes it difficult to quantitatively assess these discrepancies.

\begin{table}
    \begin{threeparttable}[b]
    \centering\small
    \caption{
        Comparison between the experimental values and the values calculated using PHITS.
        When branching ratios for both the ground and isomeric states are measured, their sum is presented.
    }
    \label{tab:PHITS}
    \begin{tabular*}{\tblwidth}{@{} CCCC @{}}
    \toprule
    Target            & Residual  & Present (\%) & PHITS (\%)  \\
    \midrule    
     $^\mathrm{nat}$Cu & $^{65}$Ni & 1.40(8)   & 3.96 \\
     $^\mathrm{nat}$Cu & $^{62}$Co & 0.618(56) & 1.18 \\ 
    \addlinespace
     $^\mathrm{nat}$Zn & $^{70}$Cu & $<$ 0.045 & 0.050 \\
     $^\mathrm{nat}$Zn & $^{69}$Cu & 0.39(9)   & 0.223 \\
     $^\mathrm{nat}$Zn & $^{68}$Cu & $>$ 1.36\tnote{1}  & 2.35 \\
     $^\mathrm{nat}$Zn & $^{67}$Cu & 11.9(11)  & 8.18 \\
     $^\mathrm{nat}$Zn & $^{66}$Cu & 9.30(47)  & 9.35 \\
    \bottomrule
    \end{tabular*}
    \begin{tablenotes}\footnotesize
        \item[]{Uncertainties include statistical and systematic uncertainties.}
        \item[1]{Lower limit, as the isomeric state of $^{68}$Cu was not measured and its contribution is not included.}
    \end{tablenotes}
    \end{threeparttable}
\end{table}

The present results are also compared with a model calculation using the Particle and Heavy Ion Transport code System (PHITS)~\cite{Sato2013-ym} version 3.33, in which muon interaction models have been implemented~\cite{Abe2017-ol}.
In this model calculation, the parameters of the surface coalescence model, which affect the emission of light complex particles, are set to $h0=0.26~\mathrm{GeV/fm}/c$ and $d=2.3$~fm, as originally proposed~\cite{Watanabe2007-cq}.
The calculated values are summarized together with the experimental values in Table~\ref{tab:PHITS}.
While the overall trends are reproduced by the PHITS calculation, quantitative discrepancies of up to a factor of a few are observed.
In particular, PHITS overestimates channels involving charged particle emission, consistent with observations reported in previous studies~\cite{Yamaguchi2025-jn}.
For the $0pxn$ channels, deviations of about a factor of a few are also observed in the PHITS calculation, and their origin is difficult to identify due to the involvement of multiple reaction channels.
These discrepancies mainly arise from the uncertainty in the initial excitation energy distribution populated by muon nuclear capture and from the limited reliability of the preequilibrium emission model~\cite{Mizuno2025-rt}.
Since the model calculation still shows a large discrepancy with the experimental result, experimentally determined branching ratios are required for beam intensity determination.

The comparison between the production BRs from muon capture on Cu and Zn has the potential to investigate the even-odd effect in the proton number. 
However, further quantitative discussion requires measurements using isotopically enriched targets. Such measurements would also enable investigations of systematic trends among Zn isotopes, as well as the origin of the large differences observed in the BRs of high-spin isomers $^{109m, 107m}$Pd produced from $^\mathrm{nat}$Ag.
The present results of the natural abundance target will be a reference for the absolute production BR of each isotope.

\subsection{Suitable target for muon beam intensity measurement} \label{sec:discussion-Imuon}

In this section, the suitability of the measured and currently available BRs for beam intensity monitoring is discussed. 
To determine the beam intensity using the in-beam activation method based on Eq.~(\ref{eq:Nmu}), the production BR and the muon nuclear capture probability ($bP_\mathrm{cap}$), and $I_\gamma$ need to be known. 
Table~\ref{tab:calib_table} summarizes the $bP_\mathrm{cap}$ values of candidate targets for absolute intensity measurement using in-beam activation, including values reported in previous measurements.
These candidates are selected based on the following criteria.
(1) $\gamma$-ray transitions with no overlap with natural backgrounds or escape peaks, ensuring reliable signal identification.
(2) Residual isotopes with negligible or no contributions from isomeric states or other parent nuclei, because the presence of multiple decay components requires solving the Bateman equations, complicating the analysis.
(3) Channels with sufficiently large BRs to ensure sufficient counting statistics.
(4) Isotopes with half-lives shorter than 10 minutes, enabling measurements within a practical time scale.
The estimated total yield of the reference peaks ($N_\gamma^\mathrm{cal}$) is also presented in Table~\ref{tab:calib_table}. 
This value is calculated with the assumption that the beam intensity of 25000 muons/second is measured for five minutes using a germanium detector with 30\% relative efficiency at 10-cm distance from the target~\cite{Mizuno2024-br}. The uncertainty includes that of $bP_\mathrm{cap}$ and the absolute uncertainty of $I_\gamma$. 
Although the uncertainty of $bP_\mathrm{cap}$ contains contributions from $I_\gamma$ and $T_{1/2}$, which are counted twice when deriving the muon beam intensity using this method, these contributions are negligible compared with the other dominant uncertainties, statistical uncertainty and the uncertainty in the detection efficiency.

Neutron emission channels are generally well suited as candidates owing to their large BRs, while several charged-particle emission channels are also feasible for light target elements when they exhibit short half-lives and high $\gamma$-ray intensities.
In practical applications, additional constraints arise from the materials used in the experimental setup.
For example, isotopes such as Al and Si can only be used when the surrounding structures are carefully designed to avoid the use of these elements, as demonstrated in previous studies.
Similarly, isotopes in the region of atomic number around 26 are subject to potential contamination from iron or stainless steel commonly used in beamline components.
Such background contributions can limit their suitability for practical applications.
From the viewpoint of target preparation, P and Mn are difficult to use owing to their fragility.
Considering the magnitude of the BR, the muon capture probability, the practical detection efficiency of the $\gamma$-ray detector, and the absence of the target material in surrounding structures, the reactions $^\mathrm{nat}$Pd ($\mu^-, \nu_\mu x$) $^{107m, 105m}$Rh and $^\mathrm{nat}$Ag ($\mu^-, \nu_\mu x$) $^{107m}$Pd are identified as ideal candidates for beam intensity measurement using the in-beam activation method.

\begin{table*}[width=1.5\linewidth,cols=6,pos=htbp]
    \centering
    \begin{threeparttable}
    \caption{Candidate reactions that can be used for the muon number calibration using the in-beam activation method. Muon nuclear capture probabilities cited in each reference paper are used for the conversion from BR to $bP_\mathrm{cap}$. 
    Information on each $\gamma$-ray is cited by NNDC, and the uncertainty of $I_\gamma$ is absolute.
    The estimation of the count of each $\gamma$-ray ($N_\gamma^\mathrm{cal}$) is also listed in the last column, based on the condition that the beam intensity of 25000 muons/second is measured for five minutes using a typical germanium detector at 10 cm. 
    }
    \label{tab:calib_table}
    \begin{tabular}{ccccccc}\toprule
    Target & Residual & $bP_\mathrm{cap}$ (\%) & $T_{1/2}$ & $E_\gamma$ (keV) & $I_\gamma$ (\%) & $N_\gamma^\mathrm{cal}$ (5 min)\\\midrule
    $^{27}$Al & $^{27}$Mg & 6.04(20)~\cite{Mizuno2025-rt} & 9.46(1) min & 843.76 & 71.80(2) & 138(5)\\
    &&&& 1014.52 & 28.80(2) & 46(2) \\
    $^{27}$Al & $^{26}$Na & 0.485(30)~\cite{Mizuno2025-rt, Yamaguchi2025-jn} & 1.07 s & 1808.71 & 99.08 & 46(3)\\
    $^{27}$Al & $^{25}$Na & 1.57(9)~\cite{Mizuno2025-rt, Yamaguchi2025-jn, Wyttenbach1978-ks} & 59.1(6) s & 389.71 & 12.6(7) & 58(4)\\
    &&&&585.03 & 13.0(7) & 41(2)\\
    &&&&974.74 & 14.9(8) & 29(2)\\
    \addlinespace
    $^\mathrm{nat}$Si & $^{27}$Mg & 1.94(8)~\cite{Mizuno2025-rt, Yamaguchi2025-jn} & 9.46(1) min & 843.76 & 71.80(2) & 44(2) \\
    &&&&1014.52 & 28.80(2) & 15(1)\\
    \addlinespace
    $^{31}$P & $^{29}$Al & 2.64(38)~\cite{Wyttenbach1978-ks} & 6.56(6) min & 1273.4 & 91.26 & 72(10)\\
    $^{31}$P & $^{28}$Al & 1.60(23)~\cite{Wyttenbach1978-ks} & 2.25 min & 1779 & 100 & 77(11)\\ 
    $^{31}$P & $^{27}$Mg & 0.90(15)~\cite{Wyttenbach1978-ks} & 9.46(1) min & 843.76 & 71.80(2) & 21(3) \\
    &&&&1014.52 & 28.80(2) & 7(1)\\
    \addlinespace
    $^{55}$Mn & $^{53}$V & 1.05(18)~\cite{Wyttenbach1978-ks} & 1.54(1) min & 1006.0 & 89.6 & 94(18)\\
    &&&&1289.1 & 10.04 & 8(2)\\
    \addlinespace
    $^{59}$Co & $^{58m}$Mn & 0.239(38)~\cite{Yamaguchi2025-jn} & 65.4(5) s & 459.2 &21.4(12) & 12(2)\\
    &&&& 810.8 & 88(5) & 30(5) \\
    &&&& 863.9 & 14.8(8) & 5(1) \\
    &&&& 1323.1 & 59(3) & 13(2)\\
    \addlinespace
    $^\mathrm{nat}$Zn & $^{69}$Cu & 0.36(8) & 2.85(15) min & (834.4)\tnote{1}\\
    &&&& 1007.5 & 23(4) & 6(1) \\
    $^\mathrm{nat}$Zn & $^{66}$Cu & 8.65(44) & 5.12(1) min & 1039.2 & 9.23 & 35(2) \\
    \addlinespace
    $^\mathrm{nat}$Pd & $^{109}$Rh  & 5.6(8)~\cite{Niikura2024-ck}  & 80.8(7) s & 326.9 & 54 & 908(299) \\
    $^\mathrm{nat}$Pd & $^{108}$Rh  & 4.2(11)~\cite{Niikura2024-ck} & 16.8(5) s & (434.1)\tnote{2} \\
    &&&& 618.9\tnote{3} & 15(4) & 152(42) \\
    $^\mathrm{nat}$Pd & $^{107m}$Rh & 5.9(4)~\cite{Niikura2024-ck}  & 0.3-10 s& 268.4 & 85.3(4)\cite{Niikura2024-ck} & 2674(182)\\
    $^\mathrm{nat}$Pd & $^{106}$Rh  & 7.8(7)~\cite{Niikura2024-ck}  & 30.1(4) s & 621.9\tnote{3} & 9.93(23) & 172(16)\\
    $^\mathrm{nat}$Pd & $^{105m}$Rh & 7.1(6)~\cite{Niikura2024-ck}  & 42.8(3) s  & 129.8 & 20.2(3) & 968(83)\\
    \addlinespace
    $^\mathrm{nat}$Ag & $^{109m}$Pd & 0.217(47) & 4.70 min & 188.9 & 56.0(3) & 26(6)\\
    $^\mathrm{nat}$Ag & $^{107m}$Pd & 1.72(10) & 21.3(5) s & 214.9 & 68.7(7) & 710(83)\\
    \addlinespace
    $^\mathrm{nat}$Ta & $^{179m}$Hf & 4.3(5)~\cite{Yamaguchi2025-jn} & 18.7 s & 160.7 & 2.88(18) & 90(12)\\
    &&&&(214.3)\tnote{4}  \\ 
    \bottomrule
    \end{tabular}
    \begin{tablenotes}\footnotesize
      \item[1] {834 keV peak overlaps with decay $\gamma$-ray originate from $^{66}$Cu.}
      \item[2] {434 keV peak overlaps with decay $\gamma$-ray originate from the isomeric state.}
      \item[3] {Several energy peaks can be found around 620 keV, and may be confusing.}
      \item[4] {214 keV peak overlaps with decay $\gamma$-ray originate from $^{178m}$Hf.}
    \end{tablenotes}
    \end{threeparttable}
\end{table*}

\subsection{Practical use for muon beam intensity measurement} \label{sec:discussion-Imuon_practical}

Using this in-beam activation measurement, the total number of muons during the measurement period can be determined using Eq.~(\ref{eq:Nmu}). Additionally, the in-beam activation method can be utilized to monitor a time-averaged beam intensity. 
When intense $\beta$-delayed $\gamma$ rays provide sufficient statistics within a given time interval, the $\gamma$-ray yield over a finite time window can be converted into the corresponding muon rate. 
For example, some of the candidates listed in Table~\ref{tab:calib_table} are expected to provide sufficient statistics within five minutes. Therefore, the $\gamma$-ray counting rate can serve as an online beam intensity monitor with a five-minute-average.
Note that the statistics shown in the last column of Table~\ref{tab:calib_table} are accumulated counts starting from the beam irradiation, and the appropriate number should be calculated depending on each timing gate during the irradiation.
Additionally, the detector dead time promptly after beam irradiation, caused by detector saturation due to the high-intensity muonic X-rays, should be corrected for the analysis.

In cases where an experiment requires beam intensity monitoring on a pulse-by-pulse basis or online $\gamma$-ray measurement is impractical, the presented method provides a useful technique for calibrating the beam intensity monitoring system. 
One example is the use of a plastic scintillator placed in the beamline, where the light output reflects the number of muons through the scintillator.
Assuming that the response of the charge integral of the plastic scintillator is linear in the region of beam intensity fluctuation, the total number of muons can be expressed as the sum of the numbers of muons in individual pulses ($n_\mathrm{beam}$) as,
\begin{equation}
    N_{\mu^-} = \Sigma_i n_\mathrm{beam}^i = \Sigma_i A\times C^i,
\end{equation}
where $C^i$ is the charge integral of the plastic scintillator for $i$-th beam pulse, and $A$ is a calibration parameter.
The calibration parameter $A$ can be determined by substituting this into Eq. (\ref{eq:Nmu}) together with the $\gamma$-ray measurement and the production BR presented in this paper.
This calibration allows subsequent measurements to be performed using only the scintillator signal.
Additionally, the calibrated plastic scintillator can be utilized to measure the intensities of positive muon beams.
Note that, since the light output of the plastic scintillator depends on the momentum of the beam, this calibration may need to be performed at several momenta and interpolated using calculations on muon energy deposit.
Another approach, which involves a stronger assumption, is to relate the muon beam intensity to the primary beam intensity monitored at the accelerator. If a proportional relationship between the primary beam and the produced muon beam is assumed, the muon intensity can be reconstructed from the accelerator beam log using the $\beta$-delayed $\gamma$-ray measurement for calibration.

When using the in-beam activation for the calibration, $r_\mathrm{decay}$ should be evaluated for each inter-pulse period using the monitored beam signal to account for beam intensity fluctuation and short beam interruptions.
To accurately account for these effects, the number of decays of the produced nuclei should be calculated for each pulse while considering the accumulation of produced nuclei during the irradiation.
By rewriting Eq.~(\ref{eq:Nmu}) in terms of the contributions from individual pulses, the total yield of the $\beta$-delayed $\gamma$-ray ($N_\gamma$) can be expressed as
\begin{equation} \label{eq:BR_finaleq}
    N_\gamma/I_\gamma\epsilon_\gamma 
    =\Sigma_{i}Y_d^i,
\end{equation}
where $\Sigma_{i=0}Y_d^i$ is the total number of decays of the residual nuclei during the measurement. 
The number of decays between the pulse number $i$-th and $i+1$-th pulses, $Y_d^i$, can be expressed as
\begin{eqnarray}\label{eq:normal_pulse_yield}
    Y_d^i&=&N_0^i \{\mathrm{exp}(-\lambda t_\mathrm{dead})-\mathrm{exp}(-\lambda T_\mathrm{gap})\}, \\
    N_0^{i}&=&n_\mathrm{beam}^{i}bP_\mathrm{cap} + N_0^{i-1} \mathrm{exp}(-\lambda T_\mathrm{gap}),
\end{eqnarray} 
where $N_0^i$ is the accumulated number of produced nuclei at the irradiation time of pulse $i$, $t_\mathrm{dead}$ is the analysis dead time after the beam irradiation, $T_\mathrm{gap}$ is the pulse distance, and $n_\mathrm{beam}^i$ is the number of muons in the single pulse. 
By relating the charge integral of the beam monitor ($C^i$) in a pulse to $n^i_\mathrm{beam}$ by $n^i_\mathrm{beam} = A \times C^i$ and using Eq.~(\ref{eq:BR_finaleq}), the calibration parameter $A$ of the beam monitor can be precisely determined.

For the systematic uncertainty in the muon beam intensity derived directly from, or calibrated using, this in-beam activation method, the uncertainty in the production BR ($b$) is generally the dominant component. 
The other nuclear quantities entering Eq.~(\ref{eq:Nmu}) or Eq.~(\ref{eq:BR_finaleq}), namely $P_{\rm cap}$, $I_\gamma$, and $T_{1/2}$ through $r_{\rm decay}$, also contribute to the uncertainty.
The contributions from $I_\gamma$ and $P_\mathrm{cap}$ depend on the selected reaction channel, although they are negligibly small in most cases.
The uncertainties in the half-life ($T_{1/2}$) are generally small, but can contribute when their magnitude is comparable to that of $b$ and the measurement time is comparable to or shorter than the half-life.
Although the systematic uncertainty depends on the reaction channels, as shown in Table~\ref{tab:calib_table}, it reaches 5--8\% for the reactions $^\mathrm{nat}$Pd ($\mu^-, \nu_\mu x$) $^{107m, 105m}$Rh and $^\mathrm{nat}$Ag ($\mu^-, \nu_\mu x$) $^{107m}$Pd, and generally ranges from 3 to 15\%.
In addition, measurement-specific uncertainties, such as those associated with the detection efficiency of each setup, should also be taken into account.

\section{Conclusion} \label{sec:conclusion}
In this study, a method for determining the absolute intensity of pulsed negative muon beams based on the in-beam activation method has been described. 
To search for suitable target materials for this method, the absolute production BRs were measured for $^\mathrm{nat}$Cu, $^\mathrm{nat}$Zn, and $^\mathrm{nat}$Ag. The measured BRs are consistent with the general trends reported in the previous studies.
The suitability of the measured and previously available production BRs for beam intensity monitoring was discussed, and their practical applications were presented.
Among the investigated reactions, $^\mathrm{nat}$Ag ($\mu^-, \nu_\mu x$) $^{107m}$Pd, with a production BR of 1.79(10)\%, was identified as a promising reference reaction for determining muon numbers in pulses owing to the absence of peak overlap with $\gamma$ rays from other residual nuclei, background radiation, and escape peaks, and the lack of any required yield correction. Together with several reactions in muonic $^\mathrm{nat}$Pd, it provides a practical reference for beam intensity measurement using the in-beam activation method.

\section*{Acknowledgment}
The muon experiment at RAL was conducted under Program No.~2070005 at the ISIS Neutron and Muon Source, allowed by the RIKEN Nishina Center Materials and Life Science PAC.
This research is partially supported by the JSPS KAKENHI Grants No.~24H00073 and No.~19H05664.
R.M. is supported by the JSPS fellows, Forefront Physics and Mathematics Program to Drive Transformation (FoPM), a World-leading Innovative Graduate Study (WINGS) program, and the JSR Fellowship from the University of Tokyo.



\bibliographystyle{elsarticle-num}

\bibliography{paperpile_man,manual}


\end{document}